\documentstyle[aps]{revtex}

\newcommand{\disregard}[1]{}

\newcommand{\Dd}{D$_{\text{2}}$}
\newcommand{\DT}{D$_{\text{2h}}^{\text{T}}$}
\newcommand{\DTD}{D$_{\text{2h}}^{\text{TD}}$}

\newcommand{\calO}{{\mathcal{O}}}

\newcommand{\hatMOmi}{\hat{\calO}^{(\mu)}}
\newcommand{\hatOmir}{\hat{O}^{(\mu)}_r}

\newcommand{\hatMKl}{\hat{\mathcal{K}}_l}

\newcommand{\hatT}{\hat{T}}

\newcommand{\hatI}{\hat{I}}

\newcommand{\barE}{\bar{E}}
\newcommand{\hatK}{\hat{K}}
\newcommand{\hatP}{\hat{P}}

\newcommand{\hatU}{\hat{U}}

\newcommand{\hatR}[1]{\hat{R}_{#1}}

\newcommand{\hatME}{\hat{\mathcal{E}}}
\newcommand{\barME}{\bar{\mathcal{E}}}
\newcommand{\hatMT}{\hat{\mathcal{T}}}
\newcommand{\barMT}{\bar{\mathcal{T}}}
\newcommand{\hatMP}{\hat{\mathcal{P}}}
\newcommand{\barMP}{\bar{\mathcal{P}}}

\newcommand{\hatMK}{\hat{\mathcal{K}}}

\newcommand{\hatMU}{\hat{\mathcal{U}}}

\newcommand{\hatMO}{\hat{\mathcal{O}}}

\newcommand{\hatMZ}{\hat{\mathcal{Z}}}

\newcommand{\hatMX}{\hat{\mathcal{X}}}

\newcommand{\hatMPT}{\hat{\mathcal{P}}^T}
\newcommand{\barMPT}{\bar{\mathcal{P}}^T}
\newcommand{\hatMRT}[1]{\hat{\mathcal{R}}_{#1}^T}
\newcommand{\barMRT}[1]{\bar{\mathcal{R}}_{#1}^T}
\newcommand{\hatMST}[1]{\hat{\mathcal{S}}_{#1}^T}
\newcommand{\barMST}[1]{\bar{\mathcal{S}}_{#1}^T}
\newcommand{\hatMR}[1]{\hat{\mathcal{R}}_{#1}}
\newcommand{\barMR}[1]{\bar{\mathcal{R}}_{#1}}
\newcommand{\hatMS}[1]{\hat{\mathcal{S}}_{#1}}
\newcommand{\barMS}[1]{\bar{\mathcal{S}}_{#1}}

\newcommand{\hatBE}{\hat{\mathbf{E}}}
\newcommand{\hatBT}{\hat{\mathbf{T}}}
\newcommand{\hatBP}{\hat{\mathbf{P}}}

\newcommand{\hatBU}{\hat{\mathbf{U}}}
\newcommand{\hatBPT}{\hat{\mathbf{P}}^T}
\newcommand{\hatBRT}[1]{\hat{\mathbf{R}}_{#1}^T}
\newcommand{\hatBST}[1]{\hat{\mathbf{S}}_{#1}^T}
\newcommand{\hatBR}[1]{\hat{\mathbf{R}}_{#1}}
\newcommand{\hatBS}[1]{\hat{\mathbf{S}}_{#1}}
\newcommand{\hatBG}[1]{\hat{\mathbf{G}}_{#1}}

\newcommand{\hatsi}{\hat{\sigma}}

\newcommand{\be}{\begin{equation}}
\newcommand{\ee}{\end{equation}}
\newcommand{\ba}{\begin{array}}
\newcommand{\ea}{\end{array}}
\newcommand{\bn}{\begin{eqnarray}}
\newcommand{\en}{\end{eqnarray}}
\newcommand{\bnl}{\begin{mathletters}\begin{eqnarray}}
\newcommand{\enl}{\end{eqnarray}\end{mathletters}}
\newcommand{\bml}{\begin{mathletters}}
\newcommand{\eml}{\end{mathletters}}
\newcommand{\bc}{\begin{center}}
\newcommand{\ec}{\end{center}}
\newcommand{\bi}{\begin{itemize}}
\newcommand{\ei}{\end{itemize}}

\newcommand{\bnll}[1]{\begin{mathletters}\label{#1}\begin{eqnarray}}
\newcommand{\enll}{\end{eqnarray}\end{mathletters}}

\newlength{\backshift}

\tighten
\begin{document}

\draft
\twocolumn[\columnwidth\textwidth\csname@twocolumnfalse\endcsname
\title{Point symmetries in the Hartree-Fock approach: Symmetry-breaking schemes}

\author{
 J. Dobaczewski,$^{1,2}$
 J. Dudek,$^{2}$
 S.G. Rohozi\'nski,$^{1}$ and
 T.R. Werner$^{1,2}$
}

\address{
$^1$Institute of Theoretical Physics, Warsaw University,
    Ho\.za 69,  PL-00681, Warsaw, Poland                            \\
$^2$Institute de Recherches Subatomiques,
    CNRS-IN$_2$P$_3$/Universit\'e
    Louis Pasteur, F-67037 Strasbourg Cedex 2, France               \\
}

\maketitle

\begin{abstract}
We analyze breaking of symmetries that belong to the double point
group {\DTD} (three mutually perpendicular symmetry axes of the
second order, inversion, and time reversal). Subgroup structure of
the {\DTD} group indicates that there can be as much as 28 physically
different, broken-symmetry mean-field schemes --- starting with
solutions obeying  all the symmetries of the {\DTD} group, through 26
generic schemes in which only a non-trivial subgroup of {\DTD} is
conserved, down to solutions that break all of the  {\DTD}
symmetries.  Choices of single-particle bases and the corresponding
structures of single-particle hermitian operators are discussed for
several subgroups of {\DTD}.
\end{abstract}

\addvspace{5mm}]

\narrowtext

\section{Introduction}
\label{sec1a}

One of the salient features of the mean-field approach to
many-fermion (e.g., nuclear) systems is the spontaneous symmetry
breaking. The symmetry of a mean-field state is called broken, if the
solution of the Hartree-Fock (HF) or Hartree-Fock-Bogolyubov (HFB)
self-consistent equations do not obey symmetries of the original
many-body Hamiltonian\cite{[RS80]}. This happens when the calculated
mean-field energy of the system is lower for states which break a
symmetry than that for unbroken symmetries. Such a mechanism depends
on the physical situation and is governed by the Jahn-Teller effect
\cite{[Jah37]}. Without going into details, let us recall that the
spontaneous breaking of an original symmetry is usually accompanied
by a significant decrease in the single-particle level density at the
Fermi energy.  Hence, the doubly magic nuclei can be safely described
by imposing conservation of the spherical symmetry, while this
symmetry should be allowed to be broken in the open-shell systems.

One of the simplest examples in this context is that of the breaking
of the translational symmetry. The related mechanism is present, e.g., in
the nuclear shell-model. Indeed,
within the framework of the shell model, interacting nucleons are
assumed to move in a common mean field that is localized in space and
consequently they cannot be described by eigenstates of the momentum
operator (plane waves). In other words, the wave functions of a
nucleus cannot be approximated by uncorrelated single-particle plane
waves - this can only be attempted for an infinite system, i.e., for the
nuclear matter. The use of a shell-model, space-localized wave
function simply reflects the correlations present in the system. In
this example, the correlations ensure that it is improbable to find
two nucleons of a nucleus at large relative distances apart.

In nuclear structure physics one can easily identify the use of
various broken symmetries in a description of well-defined,
observable effects. For instance the rotational, parity,
time-reversal, and gauge symmetry breaking were introduced to
describe the deformations, octupole correlations, nuclear rotation
and pair correlations, and combined effects thereof. At present, we
approach the situation where the mean-field calculations can be
performed without explicitly using any of the mean-field symmetries.
Several such approaches have already been implemented
\cite{[Uma91a],[Yam98],[Mat98]}, although very few calculations for
specific physical problems have been done to date.

One could, in principle, perform the mean-field calculations without
assuming {\it \`a priori} any symmetry, and let the dynamics choose
those discrete symmetries which are, in a specific situation, broken,
and those which remain obeyed. Obviously, by choosing such an
approach we cannot profit from simplifications possible when it is
known beforehand that some symmetries are obeyed/disobeyed. However,
following the general guidelines provided by the Jahn-Teller
mechanism one usually can make a reasonable choice of
obeyed/disobeyed symmetries. Such a choice is dictated by
the properties of the many-body Hamiltonian and by the classes of
phenomena which one wants to describe - it usually facilitates the
calculations markedly. In all those cases the analysis presented in
this article provides us with the mathematical means for constructing
the algorithms optimally adapted to the symmetries of the problem in
question.

In the preceding article\cite{[Dob00a]}, we have presented properties
of the single point group {\DT} and double point group {\DTD}, that
can be built from operators related to the three mutually
perpendicular symmetry axes of the second order, inversion, and time
reversal. We have also discussed their roles in the description of
even and odd fermion systems, respectively, their representations,
and the symmetry conditions induced by the conserved {\DT} or {\DTD}
symmetries on the local densities and electromagnetic moments.

By considering the {\DTD} double point group we focus on quantum
objects that are in general non-spherical, but can have one or more
symmetry axes and/or symmetry planes. Obviously, any nuclear
many-body Hamiltonian of an isolated system is time-even and
rotationally invariant. In the present paper we do not aim at
analyzing the conditions under which these symmetries are broken
spontaneously, with one or another symmetry element of the {\DTD}
group still being conserved in the HF solution. Instead, we present a
classification of all such possibilities, and discuss the resulting
properties of the mean-field Hamiltonians and single-particle wave
functions. For a review of applications of point symmetries to
a description of rotating nuclei see the recent study in
Ref.\cite{[Fra00]}.

Our goal is thus twofold:  First, in Sec.~\ref{sec3} we discuss all
possible physically meaningful subgroups of {\DTD}, and classify the
corresponding physical situations from the view-point of the
conserved {\DTD} symmetries. In many applications to date, specific
choices of conserved and broken {\DTD} symmetries have been made
\cite{[Ban69],[Gir83],[Goo85],[Naz85b],[Bon85],[Bon87],[Dob97]},
however, here we aim at a complete description of all achievable
symmetry-breaking schemes. Second, in Sec.~\ref{sec4} we review and
discuss practical aspects of structure of the mean-field operators
under specific {\DTD} group operations.  This essential question has
been explicitly or tacitly addressed in most approaches using the
deformed mean-field theory; our aim here is to present exhaustive
list of options pertaining to all the {\DTD} symmetry conditions.
Finally, conclusions are presented in Sec.~\ref{sec7a}.


\section{Subgroups of {\DT} and {\DTD} and the symmetry breaking}
\label{sec3}

The single group {\DT} and double group {\DT} \cite{[Dob00a]} can be built from
three rotations through angle
$\pi$ about the coordinate axes $k$=$x,y,z$, called the signature operators,
   \be\label{eq201}
    \hatR{k} = e^{-i\pi\hatI_k},
   \ee
to which one adds the inversion operator $\hatP$ and the time-reversal
 \be\label{eq206}
      \hatT = \bigotimes_{n=1}^A\left(-i\hatsi_y^{(n)}\right) \hatK ,
   \ee
where $\hatI_k$=$\sum_{n=1}^A{\hat{j}}_k^{(n)}$ is the total
angular momentum operator, ${\hat{j}}_k^{(n)}$ and $\hatsi_k^{(n)}$
are the angular momenta and the Pauli spin matrices for the particle
number $n$, respectively, and $\hatK$ is the complex-conjugation
operator in the coordinate representation.

{}Following the convention introduced in \cite{[Dob00a]}, with roman
symbols, like $\hatU$=$\hatR{k}$ or $\hatT$, we denote
operators acting in the Fock space
${\mathcal{H}}$$\equiv$${\mathcal{H}}_0\oplus{\mathcal{H}}_1\oplus\ldots
\oplus{\mathcal{H}}_A\oplus\ldots$. Moreover, in order to help the reader in
distinguishing between
properties of these operators when they act in even,
${\mathcal{H}}_+$$\equiv$${\mathcal{H}}_0\oplus{\mathcal{H}}_2\oplus\ldots
\oplus{\mathcal{H}}_{A=2p}\oplus\ldots$, or odd,
${\mathcal{H}}_-$$\equiv$${\mathcal{H}}_1\oplus{\mathcal{H}}_3\oplus\ldots
\oplus{\mathcal{H}}_{A=2p+1}\oplus\ldots$, fermion
spaces, we denote the former ones
with bold symbols, and the latter ones with script symbols, i.e., we
formally split the Fock-space operators $\hatU$=$\hatBU$+$\hatMU$
into two parts according to their domains.

It follows
\cite{[Kos63],[Cor84],[Dob00a]} that {\DT} is an Abelian group of 16
elements, which contains: the identity $\hatBE$, inversion $\hatBP$,
time-reversal $\hatBT$,
their product $\hatBPT$=$\hatBP\hatBT$, three signatures $\hatBR{k}$,
three simplexes $\hatBS{k}$=$\hatBP\hatBR{k}$, three $T$-signatures
$\hatBRT{k}$=$\hatBT\hatBR{k}$, and three $T$-simplexes
$\hatBST{k}$=$\hatBT\hatBS{k}$, i.e.,
   \be\label{Eq208}
      \mbox{\DT}:
      \quad
      \{ \hatBE, \hatBP,  \hatBR{k},  \hatBS{k},
         \hatBT, \hatBPT, \hatBRT{k}, \hatBST{k} \},
   \ee
where all these operators act in even-fermion-number space
${\mathcal{H}}_+$.

Similarly, the Fock-space operators $\hatU$, as well as the odd-fermion-number
operators $\hatMU$, form the group {\DTD} which is a non-Abelian group
of 32 elements. Apart from the 16 operators enumerated for {\DT},
it contains their partners obtained by multiplying every one of them by
the operators $\barE$ or $\barME$, respectively. These
operators can be identified with the rotation operators
through angle $2\pi$ about an arbitrary axis.
The partner operators are denoted by replacing the hats
with bars, i.e., the group of operators acting in ${\mathcal{H}}_-$
reads
\bn
   \mbox{\DTD} : \quad
  \{&\hatME&, \hatMP, \hatMT, \hatMPT,
        \hatMR{k}, \hatMS{k}, \hatMRT{k}, \hatMST{k},   \label{Eq219} \\
    &\barME&, \barMP, \barMT, \barMPT,
        \barMR{k}, \barMS{k}, \barMRT{k}, \barMST{k} \}.       \nonumber
\en

The complete {\DT} and {\DTD} multiplication tables have
been given and discussed in Ref.\cite{[Dob00a]}, and will not be
repeated here. We only recall a few properties of the {\DTD}
group that are essential in the following analysis, namely,
\bnll{2c01}
   \hatMR{k}^2=\hatMS{k}^2={\hatMT}^2
&=&\barME , \label{2c01e}\\
   \left({\hatMRT{k}}\right)^2=\left({\hatMST{k}}\right)^2= {\hatMP}^2
&=&\hatME , \label{2c01a}
\enll
for $k=x,y,z$,
\be
     \hatMR {k}\hatMR {l} = \hatMS {k}\hatMS {l}
   = \hatMRT{k}\hatMRT{l} = \hatMST{k}\hatMST{l}  =  \hatMR {m}, \label{2a02a}
\ee
for $(k,l,m)$ being an {\em even} permutation of $(x,y,z)$, and
\be
     \hatMR {k}\hatMR {l} = \hatMS {k}\hatMS {l}
   = \hatMRT{k}\hatMRT{l} = \hatMST{k}\hatMST{l}  =  \barMR {m}, \label{2c02a}
\ee
for $(k,l,m)$ being an {\em odd} permutation of $(x,y,z)$.

The multiplication table of {\DT} is obtained by replacing
$\barME$ and $\barMR{m}$ by $\hatBE$ and $\hatBR{m}$, respectively,
and using all bold symbols in Eqs.~(\ref{2c01})--(\ref{2c02a}).

Obviously, a
product of conserved symmetries is a conserved symmetry, and
consequently, the conserved symmetries form groups that are
subgroups of {\DT} or {\DTD}. Therefore, in order to analyze various
physically meaningful subsets of the conserved {\DT} or {\DTD}
operators, we should first consider the subgroup structure of these
groups.

Suppose that in a given physical problem, the mean-field states obey the
symmetries of a given subgroup rather than those of the whole {\DT}
or {\DTD} groups. In such a case that subgroup contains the maximal set of
operators representing the symmetry of the problem, i.e.,
all {\DT} od {\DTD} operators which do not belong to such subgroup are the
broken symmetries. From the view-point of physics,  we are
more interested in the symmetries which are broken (which is
related to interesting dynamical correlations), than in
those which are conserved. It then follows that the physically
interesting information will be attached to the operators that
{\em do not} belong to the subgroup studied, but {\em do belong}
to {\DT} or {\DTD}; those latter ones do not necessarily form a group.

{}First we consider the single group {\DT}, because (i) it is a
smaller and simpler group than {\DTD}, and (ii) the operator $\barME$
which makes the difference between the single and the double group is
always a conserved symmetry.

The analysis below is based on identifying sets of the so-called
subgroup generators, i.e, operators from which the whole given
subgroup can be obtained by their successive multiplications.
Choices of generators are, of course, non-unique, and hence in each case
we discuss and enumerate all the available possibilities.

\subsection{Subgroups of {\DT}\label{sec3a}}

Since the square of every element of {\DT} is proportional to the
identity operator $\hatBE$, we have fifteen two-element, one-generator subgroups, each of
them composed of the identity and one of the other {\DT}
operators.  We denote these subgroups by $\{\hatBG{1}\}$, where
$\hatBG{1}$ is the generic symbol corresponding to one of the
non-identity elements of {\DT}.
 Obviously, only one choice of
the generator is possible for every of the two-element
subgroups.

Similarly, group {\DT} has 35 different four-element subgroups,
which can be called the two-generator subgroups, and are denoted by symbols
$\{\hatBG{1},\hatBG{2}\}$. pertaining to their generators.
The two-generator subgroups contain, in addition to $\hatBG{1}$ and
$\hatBG{2}$, also the identity $\hatBE$ and the product
$\hatBG{1}\hatBG{2}$.  Since this product is also one of the {\DT}
operators, we have in each of the four-element, two-generator subgroups three
possibilities to select the generators.

Finally, there are 15 different eight-element, three-generator subgroups
of {\DT},
denoted by
$\{\hatBG{1},\hatBG{2},\hatBG{3}\}$.
Each of these subgroups
contains the identity $\hatBE$, the three generators,
three products of pairs of generators, and the product of
all three generators.  Hence, to choose the generators
we may first pick any pair out of seven non-identity
elements (21 possibilities), and next pick any other
subgroup element, except the product of the first two,
(4 possibilities).  Since the order in which we pick the
generators is irrelevant, one has altogether 28
possibilities of choosing the three generators in each of
the eight-element, three-generator subgroups of {\DT}.

In the same way one can calculate that there is 168 different
choices of the four generators of the whole {\DT} group;
one of them is, e.g., the set
$\{\hatBT,\hatBP,\hatBR{x},\hatBR{y}\}$.
This illustrates the degree of arbitrariness in implementing
calculations for which the whole group {\DT} is conserved.  Similar
freedom, although to a lesser degree, is available when conserving
any of the subgroups of {\DT}.  Of course, the freedom of choosing
generators cannot influence the final results, however, it
allows using different quantum numbers, phase conventions, and
structure of matrix elements, as discussed in Sec.~\ref{sec4}.

A classification of all the 65 non-trivial subgroups of {\DT}
(we do not include trivial subgroups $\{\hatBE\}$ and {\DT} itself)
is presented in
Table \ref{tab4}.  Every subgroup is assigned to a certain type,
and described by a symbol given in the first column of the Table.
The types are defined according to:  (i) the number of generators
in the subgroup (1, 2, or 3), (ii) the number of Cartesian axes
involved in the subgroup (0, I, or III standing for 0, 1, or 3),
and (iii) the number of signature operators in the subgroup (A, B,
or D standing for 0, 1, or 3).

The classification is based on two important characteristics of
each subgroup.  As shown in Ref.\cite{[Dob00a]}, every conserved
symmetry, labeled by one of the Cartesian directions $x$, $y$, or
$z$, induces a specific symmetry of local densities, related to
this particular direction.  Therefore, the number of Cartesian
axes involved in the subgroup gives us the number of symmetries of
local densities induced by the given subgroup. In addition,
the number of signature operators illustrates the way in which the
given subgroup is located with respect to the standard {\Dd}
subgroup, which is composed of the three signatures.

{\vbox{
\setlength{\backshift}{-2.5ex}
\begin{table}
\caption[TT]{
Non-trivial subgroups of the single {\DT} group,
classified according
to the types described in the text. The second column
gives the generators.
The third column gives numbers of different subgroups irrespective
of names of Cartesian axes, and
the fourth column gives the total
number of subgroups in each type.
\label{tab4}}
\begin{center}
\begin{tabular}{rl@{\hspace{\backshift}}cc}
 Type    &  \multicolumn{1}{c}{Generators}
         &                                   Generic & Total \\
\hline
 1-0$_A$: & $\{\hatBT\}$,
            $\{\hatBP\}$,
            $\{\hatBPT\}$                          &  3 &  3  \\
 1-I$_A$: & $\{\hatBRT{k}\}$,
            $\{\hatBS{k}\} $,
            $\{\hatBST{k}\}$                       &  3 &  9  \\
 1-I$_B$: & $\{\hatBR{k}\}$                        &  1 &  3  \\
\hline
          \multicolumn{2}{r@{\hspace{\backshift}}}
      {Total number of one-generator subgroups:~~~} &  7 & 15  \\
\hline
 2-0$_A$: & $\{\hatBT,\hatBP\}$                     &  1 &  1  \\
 2-I$_A$: & $\{\hatBS{k} ,\hatBT\}    $,
            $\{\hatBST{k},\hatBP\}    $,
            $\{\hatBRT{k},\hatBPT\}   $             &  3 &  9  \\
 2-I$_B$: & $\{\hatBR{k} ,\hatBT\}    $,
            $\{\hatBR{k} ,\hatBP\}    $,
            $\{\hatBR{k} ,\hatBPT\}   $             &  3 &  9  \\
2-III$_A$:& $\{\hatBS{l} ,\hatBST{m}\}$             &  1 &  6  \\
2-III$_B$:& $\{\hatBR{l} ,\hatBRT{m}\}$,
            $\{\hatBR{l} ,\hatBS{m}\} $,
            $\{\hatBR{l} ,\hatBST{m}\}$             &  3 &  9  \\
2-III$_D$:& $\{\hatBR{l} ,\hatBR{m}\} $             &  1 &  1  \\
\hline
          \multicolumn{2}{r@{\hspace{\backshift}}}
     {Total number of two-generator subgroups:~~~} & 12 & 35  \\
\hline
  3-I$_B$: & $\{\hatBR{k} ,\hatBT    ,\hatBP\}$      &  1 &  3  \\
 3-III$_B$:& $\{\hatBR{l} ,\hatBS{m} ,\hatBT\}    $,
             $\{\hatBR{l} ,\hatBST{m},\hatBP\}    $,
             $\{\hatBR{l} ,\hatBRT{m},\hatBPT\}   $  &  3 &  9  \\
 3-III$_D$:& $\{\hatBR{l} ,\hatBR{m} ,\hatBT\}    $,
             $\{\hatBR{l} ,\hatBR{m} ,\hatBP\}    $,
             $\{\hatBR{l} ,\hatBR{m} ,\hatBPT\}   $  &  3 &  3  \\
\hline
          \multicolumn{2}{r@{\hspace{\backshift}}}
    {Total number of three-generator subgroups:~~~} &  7 & 15  \\
\hline
          \multicolumn{2}{r@{\hspace{\backshift}}}
                  {Total number of subgroups:~~~} & 26 & 65  \\
\end{tabular}
\end{center}
\end{table}
}}
Classification of the subgroups of the single group {\DT} allows us
to discuss conserved and broken-symmetry schemes in even-fermion
systems.

\subsection{Subgroups of {\DTD}\label{sec3b}}

In order to discuss the conserved and broken symmetries in odd-fermion systems, we
now proceed to the discussion of the subgroups of the double group
{\DTD}. In fact, the classification of Table \ref{tab4} can now be
repeated almost without change. Indeed, whenever a given {\DT}
subgroup contains the time reversal $\hatBT$, signature $\hatBR{k}$,
or simplex $\hatBS{k}$, at least one of those, the corresponding subgroup of {\DTD}
contains  $\hatMT$, $\hatMR{k}$, or $\hatMS{k}$, and
it automatically becomes a doubled {\DTD} subgroup,
with exactly the same generators. This is so, because in the {\DTD}
group the squares of the time reversal, signature, and simplex
operators are equal to $\barME$, Eq.~(\ref{2c01e}), and hence whenever
one of these operators is present in the subgroup, it generates the
appropriate double subgroup of the double group {\DTD}. On the other
hand, when none of these generators are present in a given {\DT}
subgroup, this subgroup becomes the subgroup of {\DTD} without
doubling.

Therefore, all the {\DT} subgroups listed in Table \ref{tab4} are
simultaneously subgroups of {\DTD},
provided the generators denoted with bold symbols are replaced by the
corresponding script generators.
Most of the {\DTD} subgroups have twice
more elements than the corresponding {\DT} subgroups, with few
exceptions: subgroups $\{\hatMP\}$, $\{\hatMRT{k}\}$, $\{\hatMST{k}\}$,
$\{\hatMST{k},\hatMP\}$ do not double, and contain the same number of
elements as the corresponding subgroups of {\DT}.

{\vbox{
\setlength{\backshift}{-2.5ex}
\begin{table}
\caption[TT]{Same as in Table \protect\ref{tab4}, but for
additional subgroups of the double {\DTD} group.
\label{tab4a}}
\begin{center}
\begin{tabular}{rl@{\hspace{\backshift}}cc}
 Type    &  \multicolumn{1}{c}{Generators}
         &                                   Generic & Total \\
\hline
 2-0$_A$: & $\{\hatMP,\barME\}$                     &  1 &  1  \\
 2-I$_A$: & $\{\hatMRT{k},\barME\}$,
            $\{\hatMST{k},\barME\}$                 &  2 &  6  \\
\hline
          \multicolumn{2}{r@{\hspace{\backshift}}}
    {Total number of two-generator subgroups:~~~} &  3 &  7  \\
\hline
 3-I$_A$: & $\{\hatMST{k} ,\hatMP,\barME$\}          &  1 &  3  \\
\hline
          \multicolumn{2}{r@{\hspace{\backshift}}}
  {Total number of three-generator subgroups:~~~} &  1 &  3  \\
\hline
          \multicolumn{2}{r@{\hspace{\backshift}}}
                  {Total number of subgroups:~~~} &  4 & 10  \\
\end{tabular}
\end{center}
\end{table}
}}

In addition, these
few exceptional subgroups can be doubled explicitly by adding $\barME$
to the set of generators. For completeness, these additional subgroups
of {\DTD} are enumerated in Table \ref{tab4a}. However, the physical
contents of the additional, and of the corresponding not-doubled
subgroups from Table \ref{tab4}, are the same.
For example, they lead to exactly the
same symmetry properties of the density matrices\cite{[Dob00a]}. The
difference between them consists in the fact that the latter ones
have no irreps in the spinor space, whereas the former ones have
one-dimensional irreps with spinor bases (cf.\cite{[Kos63]}).
However, from the view-point of the symmetry breaking, they lead to
exactly the same schemes, and thus the additional subgroups shown in
Table \ref{tab4a} can be called trivial. Note, that the single-particle
operators (e.g., the mean-field Hamiltonian) are classified according
to one-dimensional representations \cite{[Dob00a]}, and hence, from
the view-point of the symmetry breaking it is irrelevant whether or
not a given subgroup has spinor representations.

Apart from the phase relations of electromagnetic moments\cite{[Dob00a]}
(note that the standard multipole operators are defined by singling out
the $z$ axis), the three Cartesian directions are, of course, entirely
equivalent.  Therefore, even though changing names of axes leads to
different subgroups of {\DT} or {\DTD}, they are identical from
the point of view of physically important features.  In Tables
\ref{tab4} and \ref{tab4a}, we give in the third columns the numbers of generic
subgroups, i.e., those which are different irrespective of
names of axes, and in the fourth column -- the total numbers of
subgroups of each type.  Index $k$ always denotes one of the
axes, i.e., $k$ can be equal to $x$, $y$, or $z$, while indices $l$
and $m$, $l$$\neq$$m$, denote one of the three pairs of different axes.

Subgroups in
types ``0'' do not depend on the Cartesian axes and,
therefore, for them the numbers of generic subgroups are equal to
the total numbers of subgroups.  Those in types ``I'' have one
generic form each, and three forms in total, depending on which
Cartesian axis is chosen.  Finally, for subgroups in types
``III'', the total numbers of subgroups can be the same, three
times larger, or six times larger (2-III$_A$) than the numbers of
generic subgroups.

In practical applications, conservation of different {\DT} or {\DTD}
subgroups may require considering either only the generic
subgroup, or all the subgroups with changed names of axes.
For example, if one considers a triaxially deformed system
with 0$^\circ$$\leq$$\gamma$$\leq$60$^\circ$, the lengths of
principal axes $a_y$$\leq$$a_x$$\leq$$a_z$ define the
orientation of the nucleus.  Then, conserved {\DT} or {\DTD} subgroups
with different names of axes may lead to different physical
consequences.  On the other hand, it can be advantageous to
consider only one generic subgroup, with a fixed orientation,
and allow for various orientations of the physical system by
extending allowed values of $\gamma$ deformation
beyond the standard first sector
0$^\circ$$\leq$$\gamma$$\leq$60$^\circ$.


\section{Single-particle bases and matrix structure of the
single-particle hermitian operators}
\label{sec4}

Throughout this section we restrict our analysis to hermitian
single-particle operators, and we study their matrix elements
in the single-particle space. Therefore, we are here concerned
with the odd number of particles (one), and hence we have to consider
the double group {\DTD}. As discussed in\cite{[Dob00a]},
the  {\DTD} operators are either linear or antilinear, and they
can have squares equal to either unity or minus unity, as
summarized in Table II of Ref.\cite{[Dob00a]}.  This gives us four categories of
operators with markedly different properties, which here will be used
for different purposes.

In the discussion which follows, we assume that the single-particle basis
is composed of pairs of time-reversed states,
and in addition we assume that the spatial wave functions
are real, i.e., not affected by the time-reversal,
\be\label{tim1}
   \hatMT|\bbox{n}\,\zeta\rangle = \zeta|\bbox{n}\,-\!\zeta\rangle,
\ee
where $\zeta$=$\pm$1 represents the intrinsic-spin degree
of freedom, and $\bbox{n}$ represents the set of
quantum numbers corresponding to space coordinates.
In particular, for the harmonic-oscillator (HO) basis,
$\bbox{n}$=$(n_x,n_y,n_z)$ are the
numbers of quanta in three Cartesian directions.
Assumption (\ref{tim1}) does not preclude whether or not the
time-reversal is a conserved operator; it only defines
the property of the single-particle basis in which
the dynamic problem is to be solved. (In principle,
the discussion below can be {\it mutatis mutandis} repeated
with $\hatMT$ replaced by $\hatMPT$, however, the use of the time-reversal
operator is more appropriate in practical applications.)

{}From now on we also assume that the basis is ordered in such a way that its
first half corresponds to the $\zeta$=$+$1 states, and the second
half is composed of their time-reversed $\zeta$=$-$1 partners.
In fact, we are entirely free to choose states in the first half
of the basis ($\zeta$=$+$1), and then Eq.~(\ref{tim1}) defines
those which belong to the second half ($\zeta$=$-$1).  In such
basis, the single-particle matrix elements corresponding to an arbitrary
hermitian operator $\hatMO$ have the form
\be\label{genmat}
   \calO = \left(\begin{array}{cc}
                            A     &   Y   \\
                       Y^\dagger  &  B
                   \end{array}
             \right)
\ee
where $A$ and $B$ are hermitian matrices, $Y$ is arbitrary,
and all submatrices are, in general,
complex.

\subsection{Single-particle bases for conserved {\DTD} operators}
\label{sec4b}

We may now separately consider several cases corresponding
to different subgroups of conserved {\DTD}
operators, Table \ref{tab4}, and to the four different
categories of operators (linear or antilinear, and hermitian or antihermitian).
In subsections \ref{sec4b1}--\ref{sec4b4}, we consider cases of various
{\DTD} generators being separately conserved, and in subsections
 \ref{sec4b5}--\ref{sec4b8}, cases of pairs of generators being simultaneously
conserved. Three-generator subgroups are briefly discussed
in Sec. \ref{sec4b9}.

\subsubsection{Time-reversal}
\label{sec4b1}

Let us first consider operators $\hatMO$ which are either even (invariant)
or odd (antiinvariant) with respect to the time-reversal:
\be\label{tim2}
    \hatMT^\dagger\hatMO \hatMT = \epsilon_T \hatMO, \qquad \epsilon_T=\pm1.
\ee
{}From Eqs. (\ref{tim1}) and (\ref{tim2}) one gets
\be\label{tim3}
   \langle \bbox{n}\,\zeta|\hatMO|\bbox{n}'\,\zeta'\rangle =
   \epsilon_T\zeta\zeta'\langle \bbox{n}\,-\!\zeta|\hatMO|\bbox{n}'\,-\!\zeta'\rangle^*.
\ee
It follows, that the matrix corresponding to $\hatMO$ reads
\be\label{matro}
   \calO = \left(\begin{array}{cc}
                            A        &       Y        \\
                     -\epsilon_T Y^* & \epsilon_T A^*
                   \end{array}
             \right),
\ee
where $A$ is hermitian, and $Y$ is antisymmetric
or symmetric for $\epsilon_T$=$+$1 and $\epsilon_T$=$-$1, respectively.
No block-diagonal structure appears, nevertheless, only two
instead of four
submatrices, $A$ and $Y$, complex in general, have to be calculated.

\subsubsection{$T$-signature or $T$-simplex}
\label{sec4b2}

As is well known\cite{[Mes62],[Dob00a]}, for each of the six antilinear {\DTD}
operators, i.e., for $T$-signature $\hatMRT{k}$ or $T$-simplex $\hatMST{k}$,
$k$=$x$,$y$,$z$, which have
squares equal to unity, ($\hatMZ^2$=$\hatME$, where $\hatMZ$ denotes one
of them), one can construct a basis composed of eigenvectors of
$\hatMZ$ with eigenvalue equal to 1,
\be\label{eq307}
   \hatMZ|\bbox{n}\,\zeta\rangle = |\bbox{n}\,\zeta\rangle.
\ee
Moreover, since every
operator $\hatMZ$ commutes with the time-reversal $\hatMT$, such basis can
always be chosen so as to fulfill condition
(\ref{tim1}) at the same time.  Table~\ref{tab-eserte} lists
examples of such bases, constructed for the HO states
$|n_xn_yn_z,s_z$=$\pm\frac{1}{2}\rangle$. A similar construction
is possible for any other single-particle basis, and has the explicit
form shown in Table~\ref{tab-eserte} provided
the space and spin degrees of freedom are separated. Note that
any linear combination of states $|\bbox{n}\,\zeta$=$+1\rangle$
and $|\bbox{n}\,\zeta$=$-1\rangle$, with real coefficients, is
another valid eigenstate of $\hatMZ$ with eigenvalue 1.

{}For operators even or odd with respect to $\hatMZ$,
\be\label{zet1}
    \hatMZ^\dagger\hatMO \hatMZ = \epsilon_Z\hatMO, \qquad \epsilon_Z=\pm1,
\ee
one then has
\be\label{zet2}
   \langle \bbox{n}\,\zeta|\hatMO|\bbox{n}'\,\zeta'\rangle =
   \epsilon_Z\langle \bbox{n}\,\zeta|\hatMO|\bbox{n}'\,\zeta'\rangle^*.
\ee
Hence, in bases fulfilling Eqs.~(\ref{tim1}) and (\ref{eq307}),
matrix $\calO$ is purely real
($\epsilon_Z$=$+$1) or purely imaginary ($\epsilon_Z$=$-$1). This gives
matrix $\calO$ in the form (tilde stands for the transposition)
\be\label{matro3}
   \left(\begin{array}{cc}
                            A        &       Y        \\
                         \widetilde{Y}      &       B
                   \end{array}
             \right)
   \qquad {\rm or} \qquad
   i\left(\begin{array}{cc}
                            A'        &       Y'        \\
                            -\widetilde{Y}'     &      -B'
                   \end{array}
             \right),
\ee
for $\epsilon_Z$=$+$1 and $\epsilon_Z$=$-$1, respectively,
where all submatrices are real, $A$ and $B$ are symmetric, $A'$ and $B'$ are
antisymmetric, and $Y$ and $Y'$ are arbitrary.
Note that in order to diagonalize $\calO$
one only needs to diagonalize a real matrix with unrestricted
eigenvalues (for $\epsilon_Z$=$+$1),
or an imaginary matrix with pairs of opposite non-zero
eigenvalues (for $\epsilon_Z$=$-$1).

{\vbox{
\begin{table}
\caption[TT]{
Examples of eigenstates $|n_xn_yn_z\,\zeta\rangle_k^T$ of
the $T$-sig\-na\-ture, $\hatMRT{k}$, or
$T$-simplex, $\hatMST{k}$ operators,
$k$=$x$, $y$, or $z$, with eigenvalue 1
[cf.\ Eqs.\ (\ref{tim1}) and (\ref{eq307})],
determined
for the harmonic oscillator states
$|n_xn_yn_z,s_z$=$\pm\frac{1}{2}\rangle$.  Symbols
$(N_x,N_y,N_z)$ refer to $(n_x,n_y,n_z)$ for $\hatMST{k}$
operators, and to $(n_y$+$n_z,n_x$+$n_z,n_x$+$n_y)$ for
$\hatMRT{k}$ operators.
\label{tab-eserte}}
\begin{center}
\begin{tabular}{lcc@{\,}r@{\,}l@{\,}c@{\,}r@{\,}l}
$k$ & $\zeta$ & \multicolumn{6}{c}{$|n_xn_yn_z\,\zeta\rangle_k^T$} \\
\hline
$x$ & $+1$
    & \multicolumn{6}{c}{$(+i)^{N_x}\exp(-i\frac{\pi}{4})\,
       |n_xn_yn_z,s_z$=$+\frac{1}{2}\rangle$}                  \\
$x$ & $-1$
    & \multicolumn{6}{c}{$(-i)^{N_x}\exp(+i\frac{\pi}{4})\,
       |n_xn_yn_z,s_z$=$-\frac{1}{2}\rangle$}                  \\[1.5ex]
$y$ & $+1$
    & \multicolumn{6}{c}{$(+i)^{N_y+1}\,
       |n_xn_yn_z,s_z$=$+\frac{1}{2}\rangle$}                  \\
$y$ & $-1$
    & \multicolumn{6}{c}{$(-i)^{N_y+1}\,
       |n_xn_yn_z,s_z$=$-\frac{1}{2}\rangle$}                  \\[1.5ex]
$z$ & $+1$ &
    & {$\frac{1}{\sqrt{2}}$}
    & ($|n_xn_yn_z,s_z$=$\frac{1}{2}\rangle $  &  $+$
    & $i(-1)^{N_z}$
    & $|n_xn_yn_z,s_z$=$-\frac{1}{2}\rangle\,)$                \\[0.5ex]
$z$ & $-1$ &
    & $\frac{i(-1)^{N_z}}{\sqrt{2}}$
    & ($|n_xn_yn_z,s_z$=$\frac{1}{2}\rangle $  &  $-$
    & {$i(-1)^{N_z}$}
    & $|n_xn_yn_z,s_z$=$-\frac{1}{2}\rangle\,)$
\end{tabular}
\end{center}
\end{table}
}}

\subsubsection{Signature or simplex}
\label{sec4b3}

Let us now consider operator $\hatMO$ which is even or odd
with respect to one of the six linear {\DTD} operators,
signatures $\hatMR{k}$ or simplexes $\hatMS{k}$,
$k$=$x$,$y$,$z$, which have
squares equal to minus unity, ($\hatMX^2$=$\barME$=$-\hatME$, where $\hatMX$ denotes
one of them), i.e.,
\be\label{tim5}
   \hatMX^\dagger\hatMO\hatMX = \epsilon_X\hatMO,
                                             \qquad \epsilon_X=\pm 1.
\ee
Since every operator $\hatMX$
commutes with the time-reversal $\hatMT$, one can always
choose a basis in which Eq.~(\ref{tim1}) holds,
and
\be\label{tim4}
   \hatMX|\bbox{n}\,\zeta\rangle = i\zeta|\bbox{n}\,\zeta\rangle,
\ee
where again $\zeta$=$\pm$1, so $r$=$\zeta{i}$=$\pm{i}$ is the signature
(for $\hatMX$=$\hatMR{k}$) or simplex $s$=$\zeta{i}$=$\pm{i}$ (for $\hatMX$=$\hatMS{k}$)
quantum number.
Table~\ref{tab-eser1} lists
such bases constructed for the HO states
$|n_xn_yn_z,s_z$=$\pm\frac{1}{2}\rangle$. A similar construction
is possible for any other single-particle basis. Note that
one can arbitrarily change the phases of states $|\bbox{n}\,\zeta$=$+1\rangle$,
and still fulfill Eqs.~(\ref{tim1}) and (\ref{tim4}); we shall use this
freedom in Secs.~\ref{sec4b8} and \ref{sec4c}.

{}From Eqs.~(\ref{tim5}) and (\ref{tim4}) one gets:
\be\label{tim6}
   \langle \bbox{n}\,\zeta|\hatMO|\bbox{n}'\,\zeta'\rangle =
   \epsilon_X\zeta\zeta'\langle \bbox{n}\,\zeta|\hatMO|
   \bbox{n}'\,\zeta'\rangle,
\ee
so the matrix $\calO$ has the form
\be\label{matfor}
   \left(\begin{array}{cc}
                   A        &       0        \\
                   0        &       B
         \end{array}
   \right)
   \qquad {\rm or} \qquad
   \left(\begin{array}{cc}
                   0        &       Y        \\
               Y^\dagger    &       0
         \end{array}
   \right)
\ee
for $\epsilon_X$=$+$1 and $\epsilon_X$=$-$1, respectively,
with $A$ and $B$ hermitian, and $Y$ arbitrary complex matrix.
Note that in
order to diagonalize $\calO$ one only needs to
diagonalize:  (i) for $\epsilon_X$=$+$1, two hermitian matrices
with complex, in general, submatrices
in twice smaller dimension, which gives real
eigenvalues with no additional restrictions,
or (ii) for $\epsilon_X$=$-$1, one hermitian
matrix $Y^\dagger Y$, again with complex submatrices in twice smaller dimension,
which gives pairs of real non-zero eigenvalues with opposite signs.

{\vbox{
\begin{table}
\caption[TT]{
Eigenstates $|n_xn_yn_z\,\zeta\rangle_k$ of the signature or
simplex operators, $\hatMR{k}$ or $\hatMS{k}$ for
$k$=$x$, $y$, or $z$, [cf.\ Eqs.\ (\ref{tim1}) and (\ref{tim4})],
determined
for the harmonic oscillator states
$|n_xn_yn_z,s_z$=$\pm\frac{1}{2}\rangle$.  Symbols
$(N_x,N_y,N_z)$ refer to $(n_x,n_y,n_z)$ for $\hatMS{k}$
operators, and to $(n_y$+$n_z,n_x$+$n_z,n_x$+$n_y)$ for
$\hatMR{k}$ operators.
Phases of eigenstates are fixed so as to fulfill condition
(\protect\ref{the-best-phases-in-the-whole-known-Universe}).
\label{tab-eser1}}
\begin{center}
\begin{tabular}{lcc@{\,}r@{\,}l@{\,}c@{\,}r@{\,}l}
$k$ & $\zeta$ & \multicolumn{6}{c}{$|n_xn_yn_z\,\zeta\rangle_k$}\\
\hline
$x$ & $+1$ &
    & $\frac{1}{\sqrt{2}}$
    & $(|n_xn_yn_z,s_z$=$\frac{1}{2}\rangle $  &  $-$
    & $(-1)^{N_x}$
    & $|n_xn_yn_z,s_z$=$-\frac{1}{2}\rangle\,)$              \\[0.5ex]
$x$ & $-1$ &
    & $\frac{(-1)^{N_x}}{\sqrt{2}}$
    & $(|n_xn_yn_z,s_z$=$\frac{1}{2}\rangle $  &  $+$
    & {$(-1)^{N_x}$}
    & $|n_xn_yn_z,s_z$=$-\frac{1}{2}\rangle\,)$              \\[1.5ex]
$y$ & $+1$ &
    & {$\frac{i^{N_y}}{\sqrt{2}}$}
    & $(|n_xn_yn_z,s_z$=$\frac{1}{2}\rangle $  &  $-$
    & $i(-1)^{N_y}$
    & $|n_xn_yn_z,s_z$=$-\frac{1}{2}\rangle\,)$              \\[0.5ex]
$y$ & $-1$ &
    & {$\frac{i^{N_y-1}}{\sqrt{2}}$}
    & $(|n_xn_yn_z,s_z$=$\frac{1}{2}\rangle $  &  $+$
    & $i(-1)^{N_y}$
    & $|n_xn_yn_z,s_z$=$-\frac{1}{2}\rangle\,)$              \\[1.5ex]
$z$ & $+1$
    & \multicolumn{6}{c}{$+i^{N_z}\exp(-i\frac{\pi}{4})\,
       |n_xn_yn_z,s_z$=$+\frac{1}{2}(-1)^{N_z+1}\rangle$}    \\
$z$ & $-1$
    & \multicolumn{6}{c}{$-i^{N_z}\exp(+i\frac{\pi}{4})\,
       |n_xn_yn_z,s_z$=$-\frac{1}{2}(-1)^{N_z+1}\rangle$}
\end{tabular}
\end{center}
\end{table}
}}

Comparing result (\ref{matfor}) with that obtained in
Sec.\ \ref{sec4b2}, one sees that the antilinear symmetries allow for
using real matrices, while linear symmetries give special
block-diagonal forms for complex matrices.

\subsubsection{Parity}
\label{sec4b4}

Standard simplification always occurs for operators which are
even with respect to the parity operator $\hatMP$,
\be\label{eq308}
   \hatMP^\dagger\hatMO\hatMP = \hatMO.
\ee
All matrices and submatrices introduced above or below
acquire a block-diagonal form, provided the single-particle bases
consist of
states with well defined parity (such as, e.g., bases listed in
Tables \ref{tab-eserte} or \ref{tab-eser1}).
Therefore, apart from Sec.~\ref{sec4b7}, we do not separately discuss
cases when the parity is one of the conserved {\DTD} operators, and we note
that the effect of the parity conservation can be easily included
on top of any other symmetry conditions.

\subsubsection{Time-reversal, and $T$-signature or $T$-simplex}
\label{sec4b5}

{}For operators $\hatMO$, for which both Eq.~(\ref{tim2}) and (\ref{zet1}) hold,
matrix $\calO$ in Eq.~(\ref{matro}) can be additionally simplified,
and reads
\be\label{matro4}
   \left(\begin{array}{cc}
                            A        &       Y        \\
                     -\epsilon_T Y   & \epsilon_T A
                   \end{array}
             \right)
   \qquad {\rm or} \qquad
   i\left(\begin{array}{cc}
                            A'        &       Y'        \\
                     \epsilon_T Y'   & -\epsilon_T A'
                   \end{array}
             \right),
\ee
for $\epsilon_Z$=$+$1 and $\epsilon_Z$=$-$1, respectively,
where all submatrices are real, $A$ is symmetric, $A'$ is
antisymmetric, $\widetilde{Y}$=$-\epsilon_T Y$, and $\widetilde{Y}'$=$-\epsilon_T Y'$.

In particular, with $\epsilon_T$=$\epsilon_Z$=$+$1, the matrix from
Eq.\ (\ref{matro4}) reduces to
\be\label{matro1}
   \calO = \left(\begin{array}{cc}
                            A        &       Y        \\
                           -Y        &       A
                   \end{array}
             \right),
\ee
where $A$ is symmetric, $Y$ antisymmetric, and both are real.
In order to diagonalize such matrix, one can consider a smaller problem,
by constructing a complex matrix $\calO_C$=\mbox{$A$$-$$iY$}
which has the size twice smaller than the
original matrix $\calO$. After diagonalizing $\calO_C$,
and separating
real and imaginary parts of its complex eigenvectors,
\be\label{matro2}
   \calO_C({u}+i{v}) =
            \omega({u}+i{v}),
\ee
one gets two degenerate real eigenvectors of $\calO$:
$\left(\begin{array}{c}
          {u}\\
          {v}
       \end{array}
 \right)$
and
$\left(\begin{array}{c}
          {v}\\
          {-u}
       \end{array}
 \right)$.
If $\hatMO$ is the mean-field Hamiltonian, such a form of eigenvectors simplifies
expressions for densities.

\subsubsection{Time-reversal, and signature or simplex}
\label{sec4b6}

{}For operators $\hatMO$, for which both Eq.~(\ref{tim2}) and (\ref{tim5}) hold,
matrix $\calO$ in Eq.~(\ref{matro}) can be additionally simplified,
and reads
\be\label{matfor2}
   \left(\begin{array}{cc}
                   A        &       0        \\
                   0        &       \epsilon_T A^*
         \end{array}
   \right)
   \qquad {\rm or} \qquad
   \left(\begin{array}{cc}
                   0        &       Y        \\
                  -\epsilon_T Y^* &       0
         \end{array}
   \right)
\ee
for $\epsilon_X$=$+$1 and $\epsilon_X$=$-$1, respectively, with
$A$ hermitian, and $Y$ antisymmetric ($\epsilon_T$=+1)
or symmetric ($\epsilon_T$=$-$1). Of course, this case is
identical to that described in Sec.~\ref{sec4b5}, because
whenever the time-reversal, and signature or simplex are conserved, the
corresponding $T$-signature or $T$-simplex are also conserved,
and $\epsilon_X$=$\epsilon_Z\epsilon_T$.
Therefore, we may now use two different bases, and obtain two different
forms of the matrix $\calO$, (\ref{matro4}) or (\ref{matfor2}),
which lead either to real, or to block-diagonal matrices.
Note that in order to diagonalize matrix $\calO$ for $\epsilon_X$=$+$1,
one has to diagonalize only its hermitian submatrix $A$, which has dimension
twice smaller than $\calO$, similarly as in Eq.~(\ref{matro2}).

\subsubsection{Parity, and signature or simplex}
\label{sec4b7}

In the {\DTD} group, the possibility of having
at ones disposal two different quantum numbers
simultaneously is very limited.  Indeed in {\DTD} one has only
three pairs of commuting linear operators, namely,
($\hatMR{k}$,$\hatMP$) for $k$=$x$, $y$, or $z$.  For each
such pair, the corresponding simplex operator $\hatMS{k}$ is also
conserved, but it does not give any additional quantum
number.  Only one generic two-generator subgroup,
$\{\hatMR{k},\hatMP\}$, see 2-I$_B$ in Table \ref{tab4},
allows, therefore, for two quantum numbers.  Similarly, only
three generic three-generator subgroups allow for two quantum numbers,
namely, (i) $\{\hatMR{k},\hatMT,\hatMP\}$, which allows only for
stationary solutions, (ii) $\{\hatMR{l},\hatMR{m},\hatMP\}$, which
does not allow for non-zero average values of the angular-momentum, and
(iii) $\{\hatMR{l},\hatMST{m},\hatMP\}$, which is the only two-quantum-number
subgroup which allows for rotating mean-field states.  Needless to say,
this latter case is most often used in cranking calculations to
date, see Sec.~\ref{sec4d}.

\subsubsection{$T$-signature or $T$-simplex, and signature or simplex}
\label{sec4b8}

Let us now consider operator $\hatMO$ which is even or odd
with respect to one of the six antilinear,
$\hatMZ^2$=$\hatME$, operators (see Sec.~\ref{sec4b2}), and
simultaneously even or odd with respect to one of
the six linear, $\hatMX^2$=$\barME$=$-\hatME$, operators (see
Sec.~\ref{sec4b3}).  In such a case, simplification of the
single-particle basis is possible only for pairs of $\hatMZ$ and
$\hatMX$ operators which correspond to two {\em different}
Cartesian directions.  Indeed, focusing our attention on
signatures, $\hatMRT{k}$ {\em commutes} with $\hatMR{k}$, and
therefore (being antilinear) it flips the $\hatMR{k}$ signature
quantum number.  Therefore, the eigenstates of $\hatMR{k}$ cannot
be eigenstates of $\hatMRT{k}$.  We may then only work either in
the basis of eigenstates of $\hatMRT{k}$, Sec.~\ref{sec4b2}, or in
the basis of eigenstates of $\hatMR{k}$, Sec.~\ref{sec4b3}.  On the
other hand, for $l$$\neq$$k$ $\hatMRT{l}$ {\em anticommutes} with
$\hatMR{k}$, and therefore, it conserves the $\hatMR{k}$ signature
quantum number.  Hence, the eigenstates of $\hatMR{k}$ can be rendered
the eigenstates of $\hatMRT{l}$ by a suitable choice of
phases.

It is easy to check that after multiplying eigenstates listed in Table
\ref{tab-eser1} by the following phase factors:
   \bnll{eq310}
   \Phi_{lk} &=& \exp\left\{i\zeta{\textstyle\frac{\pi}{2}}(N_l+1)\right\},
                             \mbox{~~~~~~~~~~~~~~~~~~for~}  k<l, \\
   \Phi_{lk} &=& \exp\left\{i\zeta{\textstyle\frac{\pi}{4}}+
                            i\zeta{\textstyle\frac{\pi}{2}}(N_l+N_k+1)\right\},
                             \mbox{~~for~}                  l<k,
   \enll
one obtains the basis states,
   \be\label{eq311}
   |\bbox{n}\,\zeta\rangle_{lk}=\Phi_{lk}|\bbox{n}\,\zeta\rangle_{k},
   \ee
which simultaneously fulfill Eqs.~(\ref{tim1}), (\ref{eq307}), and
(\ref{tim4}), i.e.,
   \bnll{eq312}
   \hatMT  |\bbox{n}\,\zeta\rangle_{lk} &=& \zeta|\bbox{n}\,-\!\zeta\rangle_{lk},\\
   \hatMZ_l|\bbox{n}\,\zeta\rangle_{lk} &=& |\bbox{n}\,\zeta\rangle_{lk}, \\
   \hatMX_k|\bbox{n}\,\zeta\rangle_{lk} &=& i\zeta|\bbox{n}\,\zeta\rangle_{lk}.
   \enll
Here, $\hatMZ_l$ stands for $\hatMRT{l}$ or $\hatMST{l}$,
  and $\hatMX_k$ stands for $\hatMR{k}$  or $\hatMS{k}$.
In Eqs.~(\ref{eq310}), symbols
$(N_x,N_y,N_z)$ refer to $(n_x,n_y,n_z)$ for the $\hatMST{l}$ or $\hatMS{k}$
operators, and to $(n_y$+$n_z,n_x$+$n_z,n_x$+$n_y)$ for the
$\hatMRT{l}$ or $\hatMR{k}$ operators. Moreover, a circular ordering
of Cartesian directions is assumed, i.e., $x$$<$$y$$<$$z$$<$$x$,
in order to define conditions $l$$<$$k$ and $k$$<$$l$.

{}For operators $\hatMO$ even or odd simultaneously with respect
to $\hatMZ_l$ and $\hatMX_k$, see Eqs.~(\ref{zet1}) and (\ref{tim5}),
bases defined by Eq.~(\ref{eq311}) allow for a very simple forms
of matrices $\calO$. Combining conditions (\ref{matro3})
and (\ref{matfor}) one obtains block diagonal {\em and} real matrix elements,
e.g. for $\epsilon_Z$=$+$1,
\be\label{matfor3}
   \left(\begin{array}{cc}
                   A        &       0        \\
                   0        &       B
         \end{array}
   \right)
   \qquad {\rm or} \qquad
   \left(\begin{array}{cc}
                   0        &       Y        \\
                  \widetilde{Y}       &       0
         \end{array}
   \right),
\ee
for $\epsilon_X$=$+$1 and $\epsilon_X$=$-$1, respectively,
with $A$ and $B$ real symmetric, and $Y$ arbitrary real matrix.

\subsubsection{Three-generator subgroups}
\label{sec4b9}

Apart from the unique case of the whole {\DTD} group being
conserved, which amounts to conserving its four generators, we also
have 15 different three-generator subgroups (Table \ref{tab4}), which
when conserved, may lead to physically different mean-field
solutions. Conserved three-generator subgroups are exceptional in
that they do not lead to further simplifications of the matrix
elements of mean-field Hamiltonians.

This is so, because cases enumerated in
Sec.~\ref{sec4b1}--\ref{sec4b8} exhaust different possibilities of
using conserved {\DTD} operators to simplify the structure of
operators by suitable choices of the single-particle bases.  Indeed,
the type ``III'' subgroups of {\DTD}, Table \ref{tab4}, which involve
operators for three different Cartesian axes, do not induce any new
simplifications. The signature or simplex
operators for different axes (the $\hatMX$ operators of
Sec.~\ref{sec4b3}) do not commute, and hence cannot give independent
quantum numbers of single-particle states.  Similarly, $T$-signature
or $T$-simplex operators for different axes (the $\hatMZ$ operators of
Sec.~\ref{sec4b2}) do not commute either, and hence cannot
simultaneously define phases of single-particle states.

One should stress, however, that even if a given conserved symmetry
does not allow for any further simplification of the matrix elements
of a mean-field Hamiltonian (like each third generator of a
three-generator subgroup), its conservation or its non-conservation
may induce entirely different solutions of the mean-field problem.

\subsection{Phase conventions}
\label{sec4c}

In Sec.~\ref{sec4b} we have shown how one can simplify the
matrix elements of operators by using a given phase convention,
i.e, by fixing phases of single-particle basis states in a given
way. Whenever an antilinear {\DTD} operator is conserved,
one can always construct a phase convention for which the matrix elements
of the mean-field Hamiltonian are real numbers. However,
from technical point of view, it can be more advantageous to
fix the phase convention in yet another way. Indeed, whenever the
calculation of matrix elements is more time consuming than
the diagonalization of the Hamiltonian matrix, one may use
the phase convention to facilitate the former task, at the expense
of diagonalizing complex matrices. Moreover, such a strategy
allows for keeping the simplicity of performing the former task
even in cases when there is no antilinear conserved symmetry
available, and when one has to diagonalize complex matrices anyhow.
In the present section
we show constructions of phase conventions which facilitate
calculations of the space-spin matrix elements.

Representation (\ref{tim1}), which separates space and spin
degrees of freedom, is convenient in applications pertaining to
deformed single-particle states, as those discussed in the
present study.  This is because, each hermitian operator can be
represented as a sum of four components of the form
\be\label{omi}
    \hatMOmi = \hatOmir\hatsi_\mu, \qquad \mu=0,1,2,3,
\ee
where $\hatOmir$ acts in the coordinate space, and $\hatsi_\mu$
are the Pauli matrices acting in the spin space, with $\hatsi_0$
defined as the 2$\times$2 identity matrix.  Then, the matrix
elements of $\hatMOmi$ can be factorized into space and spin
parts
\be\label{factor}
   \langle \bbox{n}\,\zeta|\hatMOmi|\bbox{n}'\,\zeta'\rangle =
   \langle \bbox{n}|\hatOmir|\bbox{n}'\rangle \cdot
                   \langle\zeta|\hatsi_\mu|\zeta'\rangle,
\ee
and the spin part can be computed once for all.  Usually many
of the spin matrix elements
$\langle\zeta|\hatsi_\mu|\zeta'\rangle$ vanish, thus making
it unnecessary to calculate the corresponding
coordinate-space matrix elements $\langle
\bbox{n}|\hatOmir|\bbox{n}'\rangle$.

Matrix elements of operators $\hatMOmi$ can be made purely real
or purely imaginary if phases of single-particle basis states are
chosen in such a way that, for one Cartesian direction
$l$=$x$, $y$, or $z$, one has
\be\label{kazet}
   \hatMKl|\bbox{n}\,\zeta\rangle = |\bbox{n}\,\zeta\rangle,
\ee
where $\hatMKl^2$=$\hatME$ is the antilinear spin operator
defined by
\be\label{kaka}
   \hatMKl = \hatMT i\hatsi_l = i\hatsi_l \hatMT .
\ee
Indeed, for time-even ($\epsilon_T$=$+$1) or time-odd
($\epsilon_T$=$-$1) operators one obtains that
\be\label{kztrans}
   \hatMKl^\dagger\hatMOmi\hatMKl
       = \epsilon_{\mu l}\hatMT^\dagger\hatMOmi\hatMT
       = \epsilon_{\mu l}\epsilon_T\hatMOmi,
                                        \quad \epsilon_{\mu l}=\pm 1,
\ee
where coefficients $\epsilon_{\mu l}$ are given
in Table~\ref{tab-eser}.
Using Eqs.~(\ref{kazet}) and (\ref{kztrans}), one gets for matrix
elements of $\hatMOmi$
\bnll{nzme}
   \langle \bbox{n}\,\zeta|\hatMOmi|\bbox{n}'\,\zeta'\rangle
       &=& \epsilon_{\mu l}\epsilon_T
           \langle \bbox{n}\,\zeta|\hatMOmi|\bbox{n}'\,\zeta'\rangle^*
                                                        \label{nzmeb}\\
   \langle \bbox{n}\,\zeta|\hatMOmi|\bbox{n}'\,\zeta'\rangle
       &=& \epsilon_{\mu l}\zeta\zeta'
           \langle \bbox{n}\,-\!\zeta|\hatMOmi|\bbox{n}'\,-\!\zeta'\rangle,
                                                        \label{nzmec}
\enll
where (\ref{nzmeb}) tells us which elements are real, and which are
imaginary, while (\ref{nzmec}) gives the matrix elements, e.g., for
$\zeta$=$-$1 expressed through those for $\zeta$=$+$1.

As is usual for antilinear operators, there is
a lot of freedom in finding bases (\ref{kazet}) of eigenstates of $\hatMKl$.
We can use this freedom to fulfill other useful conditions.
For example, since $\hatMKl$ and $\hatMR{k}$ anticommute for
$l$$\neq$$k$, one can find bases (\ref{kazet}) which are at the
same time the eigenstates of signature or simplex operators. In fact,
phases of eigenstates listed in Table \ref{tab-eser1}, has been
chosen in such a way that,
\be\label{the-best-phases-in-the-whole-known-Universe}
    \hatMKl|\bbox{n}\,\zeta\rangle_k = |\bbox{n}\,\zeta\rangle_k,
     \mbox{~~for~~} k<l,
\ee
where again the circular ordering of Cartesian directions,
$x$$<$$y$$<$$z$$<$$x$,
is assumed to define $k$$<$$l$.

{\vbox{
\begin{table}
\caption[TT]{
Antilinear spin operators $\hatMKl$, which can be used to fix convenient
phase conventions leading to conditions (\protect\ref{nzme}).
\label{tab-eser}}
\begin{center}
\begin{tabular}{lr@{=}lcccc}
$l$ & \multicolumn{2}{c}{$\hatMKl$}
    &           $\epsilon_{0l}$ & $\epsilon_{1l}$ &
                $\epsilon_{2l}$ & $\epsilon_{3l}$                 \\
\hline
$x$ &  ${\hatMT}i\hatsi_x$ & $i\hatsi_z\hatMK$  & $+1$ & $+1$ & $-1$ & $-1$ \\
$y$ &  ${\hatMT}i\hatsi_y$ & $\hatMK$           & $+1$ & $-1$ & $+1$ & $-1$ \\
$z$ &  ${\hatMT}i\hatsi_z$ & $-i\hatsi_x\hatMK$ & $+1$ & $-1$ & $-1$ & $+1$ \\
\end{tabular}
\end{center}
\end{table}
}}

\subsection{Examples of previous cranking approaches}
\label{sec4d}

As argued in Sec.~\ref{sec4b7}, there are good reasons to use in
cranking calculations the subgroup $\{\hatMR{l},\hatMST{m},\hatMP\}$
(Table \ref{tab4}) of conserved {\DTD} symmetries. This generic
three-generator subgroup appears in three space orientations, i.e.,
for $l$=$x$, $y$, or $z$, and each of these possibilities was
employed in one of the HF(B) or phenomenological-mean-field cranking
analyses to date.

In particular, traditionally the $x$ axis was chosen as
the direction of the cranking angular momentum, see
e.g.~Ref.~\cite{[Naz85b]}, and therefore, the standard Goodman
basis \cite{[Goo85]} corresponds to the $l$=$x$ subgroup, with
phases of single-particle states (and quasiparticle states,
for that matter) fixed by using the $\hatMRT{z}$ operator.
Then, by dropping the parity operator from the symmetry group
$\{\hatMR{x},\hatMST{z},\hatMP\}$, most octupole-cranking
calculations were performed within the 2-III$_A$ subgroup
$\{\hatMS{x},\hatMST{y}\}$ of Table \ref{tab4}.

Another choice was made in the HO-basis \cite{[Gir83]} and
coordinate-space \cite{[Bon85],[Bon87]} HF(B) calculations,
where the $z$ axis was used as the cranking axis.  Such choice
was motivated by the standard representation of spinors,
that are eigenstates of $\hatsi_z$, and hence the $l$=$z$
subgroup $\{\hatMR{z},\hatMST{y},\hatMP\}$ was employed.
In these approaches, phases of
single-particle states were fixed by using the $\hatMST{y}$
operator, and the parity-broken calculations were done within
the $\{\hatMS{z},\hatMST{y}\}$ subgroup.

Finally, in the recent Cartesian HO-basis HF approach of
Ref.~\cite{[Dob97]}, the code HFODD was constructed for the
conserved $l$=$y$ subgroup $\{\hatMR{y},\hatMST{x},\hatMP\}$,
and the $y$ direction was used for the
cranking axis.
The choice of this symmetry, and the resulting choice of the
$y$ cranking axis, was motivated by the fact that it allows for
using real electric multipole moments, cf.~Ref.\cite{[Dob00a]}.
Phases of single-particle states were in Ref.~\cite{[Dob97]} fixed
by using the $\hatMK_z$ operator (\ref{kaka}), and
calculations were performed within the basis of the
$\hatMS{y}$ eigenstates, Table \ref{tab-eser1}.  The HFODD
code allows for calculations with
one symmetry plane, and this is done within the
$\{\hatMS{y}\}$ conserved symmetry group of Table \ref{tab4}.
The code can also optionally perform the
two-symmetry-plane cranking calculations for the 2-III$_A$
subgroups $\{\hatMS{y},\hatMST{x}\}$ and
$\{\hatMS{y},\hatMST{z}\}$.

\section{Conclusions}
\label{sec7a}

We have analyzed the "far end" of the symmetry breaking chain,
namely, symmetries of mean-field nuclear states, which range from
time-even, parity-even, signature conserving states, (nevertheless
breaking the rotational and axial symmetry), to those which do not
conserve any symmetries at all.  We have shown that intermediate
cases, between such two extremes, correspond to conserved subgroups
of the {\DT} or {\DTD} point symmetry groups.  A classification of all the subgroups
has been proposed, and we have shown that there are 26 different
non-trivial symmetry-breaking schemes, when names of Cartesian axes
are irrelevant, and 65 different non-trivial symmetry-breaking
schemes when names of axes are distinguished in the intrinsic frame
of reference.

Consequences of conserving individual {\DTD} symmetries have been
enumerated for the construction of single-particle bases in which
mean-field operators may have special simplified forms. We point out
that the same forms of the mean-field Hamiltonian may correspond to
different conserved symmetries, and hence to different physical
consequences for observables obtained in the mean-field methods. We
have also analyzed and compared various options for defining phase
conventions of single-particle basis states.

\acknowledgments

This research was supported in part by the Polish Committee for
Scientific Research (KBN) under Contract Nos.~2~P03B~034~08 and
2~P03B~040~14, and by the French-Polish integrated actions
programme POLONIUM.


\end{document}